\documentclass[runningheads]{CMSIM}

\usepackage{graphicx} 
\usepackage{amsmath,amssymb}                      

\renewcommand{\thefootnote}

\title*{Stochastic Evolution of Stock Market Volume-Price Distributions}

\titlerunning{Stochastic Evolution of Stock Market Distributions}

\author{
  Paulo Rocha\inst{1}, 
  Frank Raischel\inst{2}, 
  Jo\~ao P.~da Cruz\inst{3,4}, 
  and Pedro G.~Lind\inst{4,5}
}

\authorrunning{\it Rocha et al.}

\institute{%
  Mathematical Department, FCUL
  University of Lisbon,
  1749-016 Lisbon, Portugal\\
  (e-mail: {\tt paulorocha99@hotmail.com})
  \and
  Instituto Dom Luiz, CGUL,
  University of Lisbon,
  1749-016 Lisbon, Portugal\\
  (e-mail: {\tt raischel@cii.fc.ul.pt})
  \and
  Closer Consulting LTD,
  4-6 University Way,
  London E16-2RD, United Kingdom\\
  (e-mail: {\tt joao.cruz@closer.pt})
  \and
  Centro F\'{\i}sica Te\'orica e Computacional,
  Avenida Prof.~Gama Pinto 2,
  1649-003 Lisboa, Portugal\\
  (e-mail: {\tt joao.cruz@closer.pt})
  \and
  ForWind and Institute of Physics,
  University of Oldenburg,
  DE-26111 Oldenburg, Germany\\
  (e-mail: {\tt pedro.g.lind@forwind.de})
}

\begin{document}
\thispagestyle{empty}
\maketitle             
\setlength{\leftskip}{0pt}
\setlength{\headsep}{16pt}
\footnote{\begin{tabular}{p{11.2cm}r}
\small {\it $3^{rd}$SMTDA Conference Proceedings, 11-14 June 2014, Lisbon Portugal } \\  
 \small C. H. Skiadas (Ed)\\  \small \textcopyright {} 2014 ISAST & \includegraphics[scale=0.38]{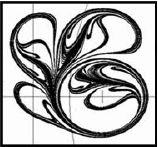}
 \end{tabular}}

\begin{abstract}
Using available data from the New York stock market (NYSM)
we test four different biparametric models to fit the 
correspondent volume-price distributions at each $10$-minute lag: the 
Gamma distribution, the inverse Gamma distribution, the Weibull 
distribution and the log-normal distribution.
The volume-price data, which measures market capitalization,  
appears to follow a specific statistical pattern, other than the 
evolution of prices measured in similar studies. 
We find  that the inverse Gamma model gives a superior fit to 
the volume-price evolution than the other models. 
We then focus on the inverse Gamma distribution as a model for
the NYSM data and analyse the evolution of its
distribution parameters as a stochastic process.
Assuming that the evolution of these parameters is governed by 
coupled Langevin equations, we derive the corresponding 
drift and diffusion coefficients, which then provide 
insight for understanding the mechanisms underlying the evolution
of the stock market.
\end{abstract}


\section{Scope and Motivation}

In 1973 a breakthrough in financial modelling was proposed by Black and
Scholes, who reinterpreted the Langevin equation for Brownian motion to 
predict value European options, assuming the underlying asset follows a 
stochastic process in the form\cite{bsmodel1,bsmodel2} 
\begin{equation}
\frac{dS_t}{S_t}=\mu dt+\sigma d W_t, 
\label{GBM}
\end{equation}
for $S_0>0$, where $S_t$ is the asset price, $\mu$ is the mean rate of 
the asset return and $W_t$ describes a Wiener process, 
with distribution $W_t\sim N(0,t)$.
The value of $\sigma$, so-called volatility, measures the risk associated 
to the fluctuation of the asset return. Thus, by making a good estimate
of its value one is able to establish a criterion for selling and buying 
in order to optimize the profit.

The BS, and similar stochastic approaches based on Gaussian uncorrelated 
noise sources, have since then received both strong criticism and 
improvements, such as  stochastic volatility models\cite{Heston}. 
It has been acknowledged that in more realistic models the statistics 
of extreme events, leading to heavy tails in the distributions, as
well as correlations between noise sources and other components need to 
be taken into account.


In this paper we put this important extension in a more general
context.
From a purely mathematical perspective, for each stochastic variable
obeying a given Langevin equation there is a probability 
density function (PDF) associated to it that fulfils a Fokker-Planck
equation\cite{Risken}. Probability density functions are defined
by a few parameters that characterize the corresponding statistical
moments. The generalization of the Black-Scholes model to incorporate 
stochastic volatility is a particular case of having one probability 
density function whose parameters are themselves stochastic variables 
governed by stochastic differential equations.
By modelling such ``stochastic'' probability density functions one is
able to properly describe how they evolve and, thus, evaluate how uncertain 
is a given prediction of the corresponding variable. 
We focus here on the evolution of the volume-price, i.e.~on changes in 
capitalization, which should have more the character of a conserved 
quantity than the price per se. While the price and volume distribution
are useful for portfolio purposes, to have access to the overall distribution
of volume-prices provides information about the entire capital traded in
the market.

In this paper, we show that heavy tails are present in the statistics of 
the capitalization, and we specifically present a stochastic evolution 
equation for the tail parameter. 
In the context of finance models, such approach can eventually enable one 
to improve measures of risk and to provide additional insight in risk 
management.


We start in Sec.~\ref{sec:data} by describing the data collected
from the New York stock market and in Sec.~\ref{sec:models} we
apply four typical models in finance to fit the empirical
data. We will argue that inverse Gamma is a good model for the
cumulative distributions of volume-prices and therefore, in 
Sec.~\ref{sec:inverseGamma}, we concentrate in its fit parameters 
to mathematically describe the stochastic evolution of volume-price
distributions. Conclusions close the paper in Sec.~\ref{sec:conclusions}.
\begin{figure}[t]
\centering
\includegraphics[width=0.95\textwidth]{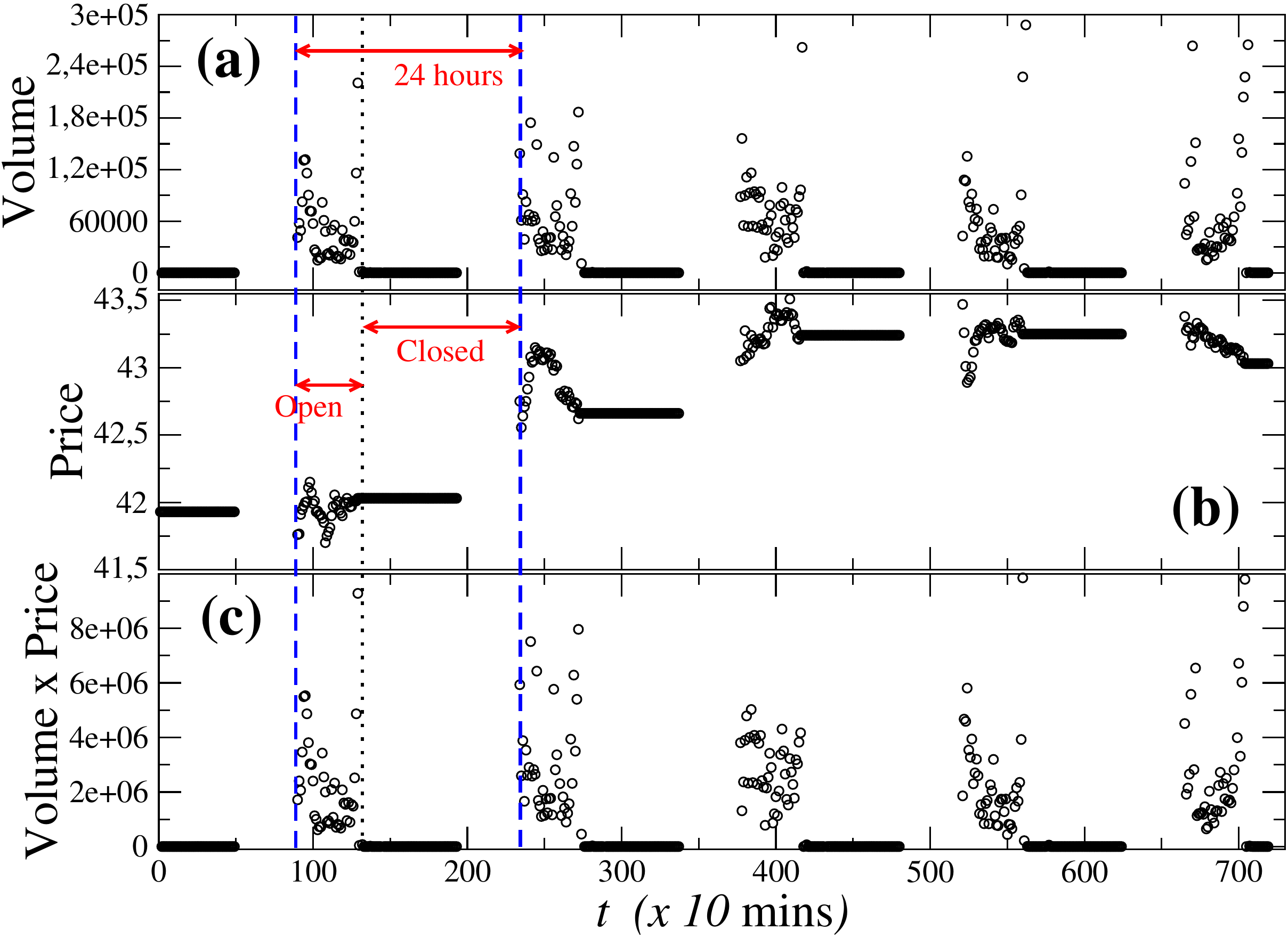}
\caption{\protect
         Illustration of the volume and price evolution for one 
         company during four days:
         {\bf (a)} volume $V$, 
         {\bf (b)} price $p$ and
         {\bf (c)} volume-price $pV$ time-series.}
\label{fig01}
\end{figure}
\begin{figure}[htb]
\centering
\includegraphics[width=0.9\textwidth]{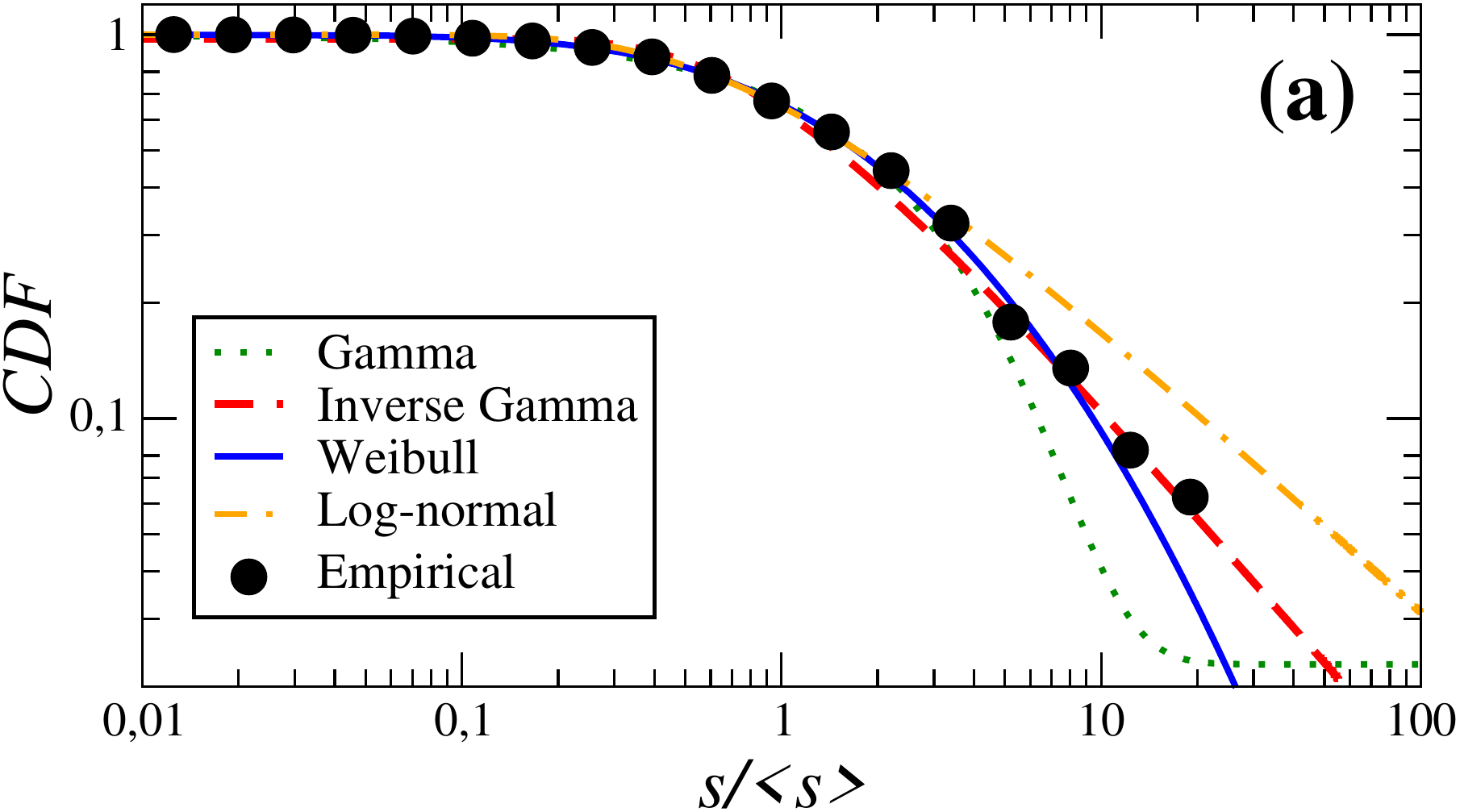}
\includegraphics[width=0.9\textwidth]{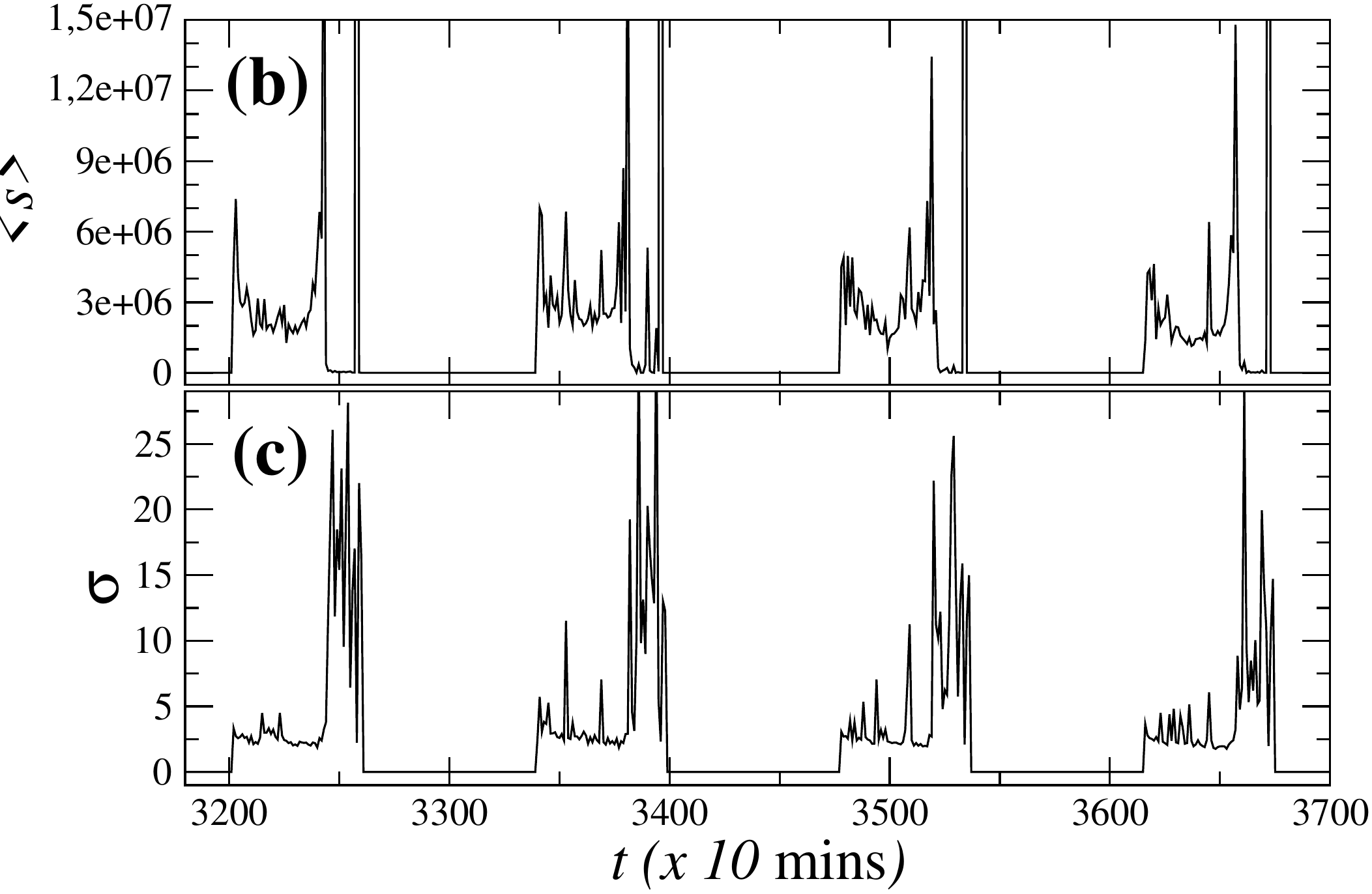}
\caption{\protect
         {\bf (a)}
         Numerical cumulative density function fitted by the 
         four different distributions:
         log-normal distribution 
         $\Gamma-$distribution,
         inverse $\Gamma-$distribution,
         Weibull-distribution.
         To characterize the evolution of the density functions
         one first considers the time series of the 
         {\bf (a)} empirical volume-price average $\langle s\rangle$ and of the 
         {\bf (b)} corresponding standard deviation $\sigma$.}
\label{fig02}
\end{figure}
\begin{figure}[htb]
\centering
\includegraphics[width=0.95\textwidth]{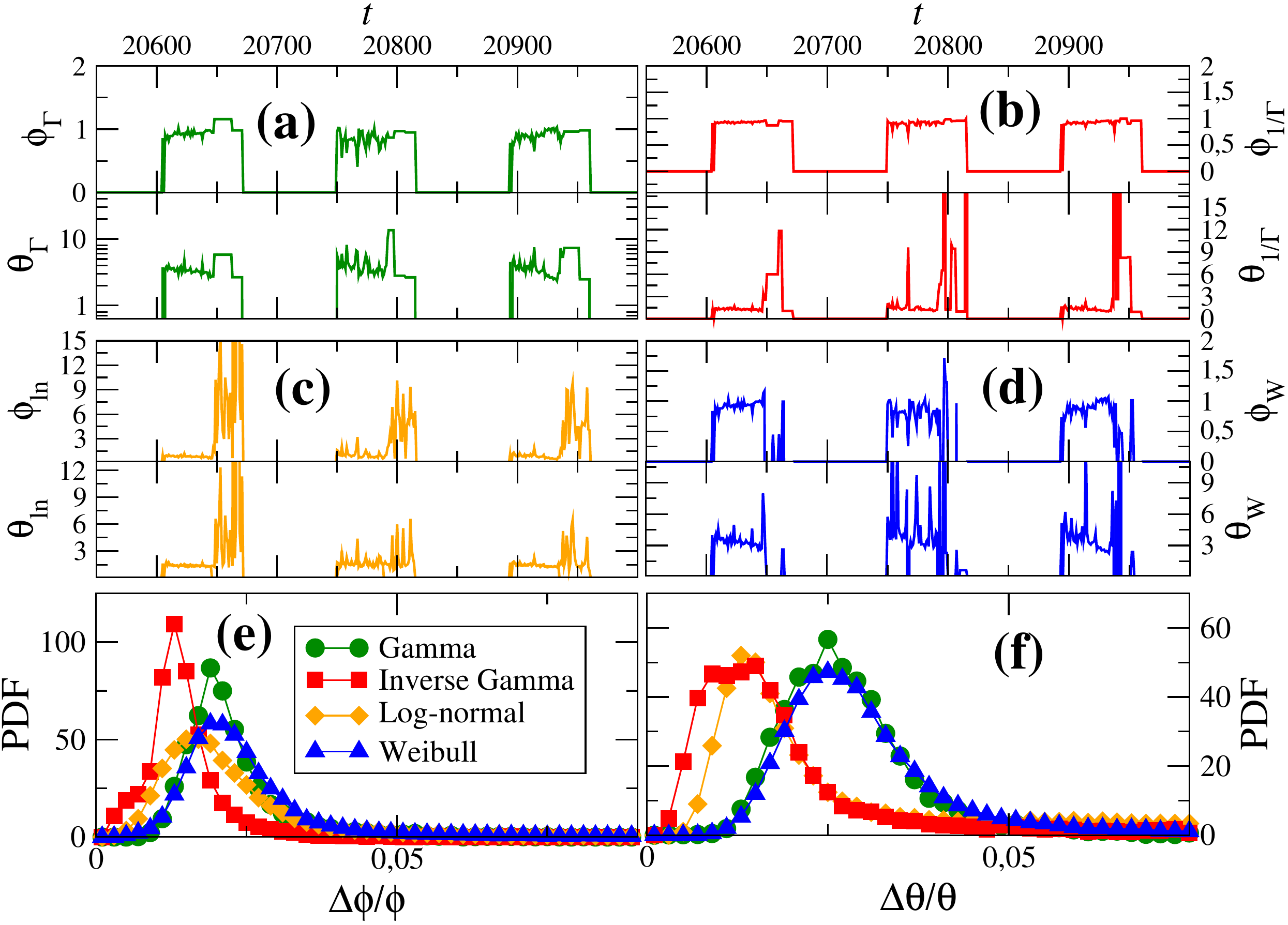}
\caption{\protect
         Time series of the two parameters characterizing the evolution of 
         the cumulative density function (CDF) of the volume-price $s$:
         {\bf (a)} $\Gamma$-distribution
         {\bf (b)} inverse $\Gamma$-distribution,
         {\bf (c)} log-normal distribution and
         {\bf (d)} Weibull distribution.
         Each point in these time series correspond to $10$-minute intervals. 
         Periods with no activity correspond to the period where market 
         is closed, and therefore will not be considered in our approach.
         {\bf (e-f)} Probability density function of the resulting relative 
         error correspondent to the fitting parameters $\phi$ and $\theta$ 
         for each distribution. In all plots, different colors correspond
         to different distribution models.}
\label{fig03}
\end{figure}

\section{Data}
\label{sec:data}

We construct a database of several listed shares extracted 
from the New York stock market (NYSM) every ten minutes starting in March 16th, 
2011 to January 1st, 2014. From the data, we compute volumes distributions 
for each ten minutes, in order to obtain a full description of the temporal 
evolution of the transactions.     
All the data were collected from the website {\tt http:}\-{\tt //finance.}\-{\tt yahoo.com/} 
every $10$ minutes during almost three years ($907$ days), yielding a 
total of $N_p\sim 10^5$ data points. 

Each register refers to one specific listed company and is composed
by the following fields: last trade price, volume, day's high price, 
day's low price, last trade date, 200 days-moving average, average daily 
volume and company name. 
In total, we were able to have a total of $N_e\sim 2000$ 
listed companies for each time-span of $10$ minutes.
Since we do not have access to the instantaneous trading price of each 
transaction for each company, we consider the last trade price as the 
estimate of the price change on each set of ten minutes trading volume.

Figure \ref{fig01}a and \ref{fig01}b show the evolution of the trading 
volume $V$ and the last trade price $p$ respectively for one single
company during approximately 5 working days.
We define the volume-price $s=pV$ as the product of both these 
properties (see Fig.~\ref{fig01}c) and will concentrate henceforth 
in analysing its joint evolution.
This image gives us an idea of how our volume-price $s$ and 
the separated components, volume $V$ and price $p$, change 
along one day in one particular company and, consequently, it 
reflects the 
change in capitalization of a given company.

In Fig.~\ref{fig01} we also indicate that the period of six and 
half hours during which the price change, corresponds exactly to the 
period at which the NYSM is open, generally from 9:30 am to 4:00 pm
(east time).
After the market closes, there is still a $4$-hour window during which
trading occurs, so-called after-hours trading, typically from 
4:00 to 8:00 pm.
We maintain these largely inactive periods for future studies on the 
statistics of the after-hours trade. 
In the context of this study, the changes in 
capitalization during these periods can be neglected.

For each $10$-minute interval we compute the cumulative
density distribution (CDF) of all $N_e$ volume-prices
and record its respective average $\langle s\rangle$
over the listed companies,
and standard deviation $\sigma$.
For convenience, we take the volume-price normalized to
its average $\langle s \rangle$ when computing the CDF.
In Fig.~\ref{fig02}a we show the CDF for a particular
$10$-minute span and in Fig.~\ref{fig02}b and \ref{fig02}c
one plots the typical evolution of the average and standard
deviation respectively.
\begin{table}[b]  
\centering
\resizebox{\columnwidth}{!}{
\begin{tabular}{l|cc|cc}
\hline
\\[-2.0ex]
 & \multicolumn{2}{c|}{Param.~err.~$\Delta\phi/\phi$} & \multicolumn{2}{c}{Param.~err.~$\Delta\theta/\theta$} 
\\[-0.1ex]
\hline 
&\raisebox{-1.0ex}{Average} & \raisebox{-1.0ex}{Std Dev.} &\raisebox{-1.0ex}{Average} & \raisebox{-1.0ex}{Std Dev.}  \\[1ex]
\hline
\raisebox{-1.0ex}{$\Gamma-$distribution} & 
\raisebox{-1.0ex}{2.21e-2} & 
\raisebox{-1.0ex}{8.54e-3} &
\raisebox{-1.0ex}{2.82e-2} &
\raisebox{-1.0ex}{1.16e-2}\\[1ex]
\raisebox{-1.0ex}{{\bf Inverse $\Gamma-$distribution}} & 
\raisebox{-1.0ex}{{\bf 1.43e-2}} & 
\raisebox{-1.0ex}{{\bf 6.46e-3}} &
\raisebox{-1.0ex}{{\bf 3.43e-2}} &
\raisebox{-1.0ex}{{\bf 5.49e-2}}
\\[1ex]
\raisebox{-1.0ex}{Weibull} & 
\raisebox{-1.0ex}{3.13e-2} & 
\raisebox{-1.0ex}{5.29e-2} &
\raisebox{-1.0ex}{4.89e-2} &
\raisebox{-1.0ex}{9.77e-2}\\[1ex]
\raisebox{-1.0ex}{Log-normal} & 
\raisebox{-1.0ex}{3.78e-2} & 
\raisebox{-1.0ex}{7.53e-2} &
\raisebox{-1.0ex}{5.60e-2} &
\raisebox{-1.0ex}{9.28e-2}\\[1ex]
\hline 
\end{tabular}
}
\caption{\protect
         The average and standard deviations of the value distributions
         for each parameter error, $\Delta\phi/\phi$ and $\Delta\theta/\theta$,
         in Fig.~\ref{fig03}e-f. The best fit is indeed obtained for
         the inverse Gamma distribution.}
\label{tab1}
\end{table}

The choice of the normalized volume-price is the best for
assessing the underlying ``geometry'' of the market as a 
complex network\cite{ourpaper3}, and therefore we consider
henceforth the normalized volume-price $s/\langle s\rangle$.
Volume-price represents the amount of capital of a particular 
listed company that is exchanged in the market. 
The normalized distribution of volume-price represents the 
distribution of links between investors and companies.

\section{Four models for volume-price distributions}
\label{sec:models}

In order to find a good fit to the empirical CDF 
we will consider four well-known bi-parametric distributions, namely the
Gamma distribution, inverse Gamma distribution, log-normal distribution
and the Weibull distribution.
We fit the empirical CDF data (bullets in Fig.~\ref{fig02}a)
with these four different models, which are often used for
finance data analysis\cite{silvio13}. 

The Gamma probability density function (PDF) is given by
\begin{equation}
F_{\Gamma}(s)= \frac{s^{\phi_{\Gamma}-1}}{\theta_{\Gamma}^{\phi_{\Gamma}}\Gamma[\phi_{\Gamma}]}exp\left[-\frac{s}{\theta_{\Gamma}}\right] ,
\label{Gamma-distribution_PDF}
\end{equation}
the inverse Gamma PDF by
\begin{equation}
F_{1/\Gamma}(s)= \frac{\theta_{1/\Gamma}^{\phi_{1/\Gamma}}}{\Gamma[\phi_{1/\Gamma}]}s^{-\phi_{1/\Gamma}-1}exp\left[-\frac{\theta_{1/\Gamma}}{s}\right] ,
\label{Inverse_Gamma-distribution_PDF}
\end{equation}
the log-normal PDF by
\begin{equation}
F_{\hbox{ln}}(s)= \frac{1}{s\theta_{\hbox{ln}}\sqrt{2\pi}}exp\left[-\frac{(\log s-\phi_{\hbox{ln}})^2}{2\theta_{\hbox{ln}}^2}\right]
\label{Log-normal_PDF}
\end{equation}
and the Weibull PDF by
\begin{equation}
F_{W}(s)= \frac{\phi_W}{\theta^{\phi_W}_W}s^{\phi_W-1}exp\left[-\left(\frac{s}{\theta_W}\right)^{\phi_W}\right] .
\label{Weibull-distribution_PDF}
\end{equation}

In Fig.~\ref{fig02}a we plot the corresponding fit of each of these models
for the empirical CDF.
In Fig.~\ref{fig03}(a-d) we show a short time-interval of the series
of each pair of parameter.

For each model above, we take into account the relative error of each 
parameter value, $\Delta\phi/\phi$ and $\Delta\theta/\theta$, computed using 
a least square scheme when making the fit.
Figure \ref{fig03}e and \ref{fig03}f show the distributions of the 
observed relative errors of $\phi$ and $\theta$ respectively.
From these two plots it seems that each distribution fits quite well 
the empirical CDF data, since
relative errors are mostly under five percent.
From the inspection of Fig.~\ref{fig03}e and \ref{fig03}f as well
as Tab.~\ref{tab1}, one sees
that the best fit seems to be for the inverse Gamma distribution and
therefore we will consider henceforth only this distribution. 

\section{The stochastic evolution of inverse Gamma tails}
\label{sec:inverseGamma}

To explore the inverse Gamma distribution model, we first consider
the meaning of its two parameters. 
A closer look at Eq.~(\ref{Inverse_Gamma-distribution_PDF}) leads to the
conclusion that while $\theta$ characterizes the shape of the
distribution for the lowest range of volume-prices, the parameter
$\phi$ characterizes the power law tail $\sim s^{-\phi-1}$.
Since it is this tail that incorporates the large fluctuations 
of volume-prices, in this section we focus on the evolution of the
parameter $\phi$ solely. Label $1/\Gamma$ is dropped for
simplicity. 
\begin{figure}[t]
\centering
\includegraphics[width=0.95\textwidth]{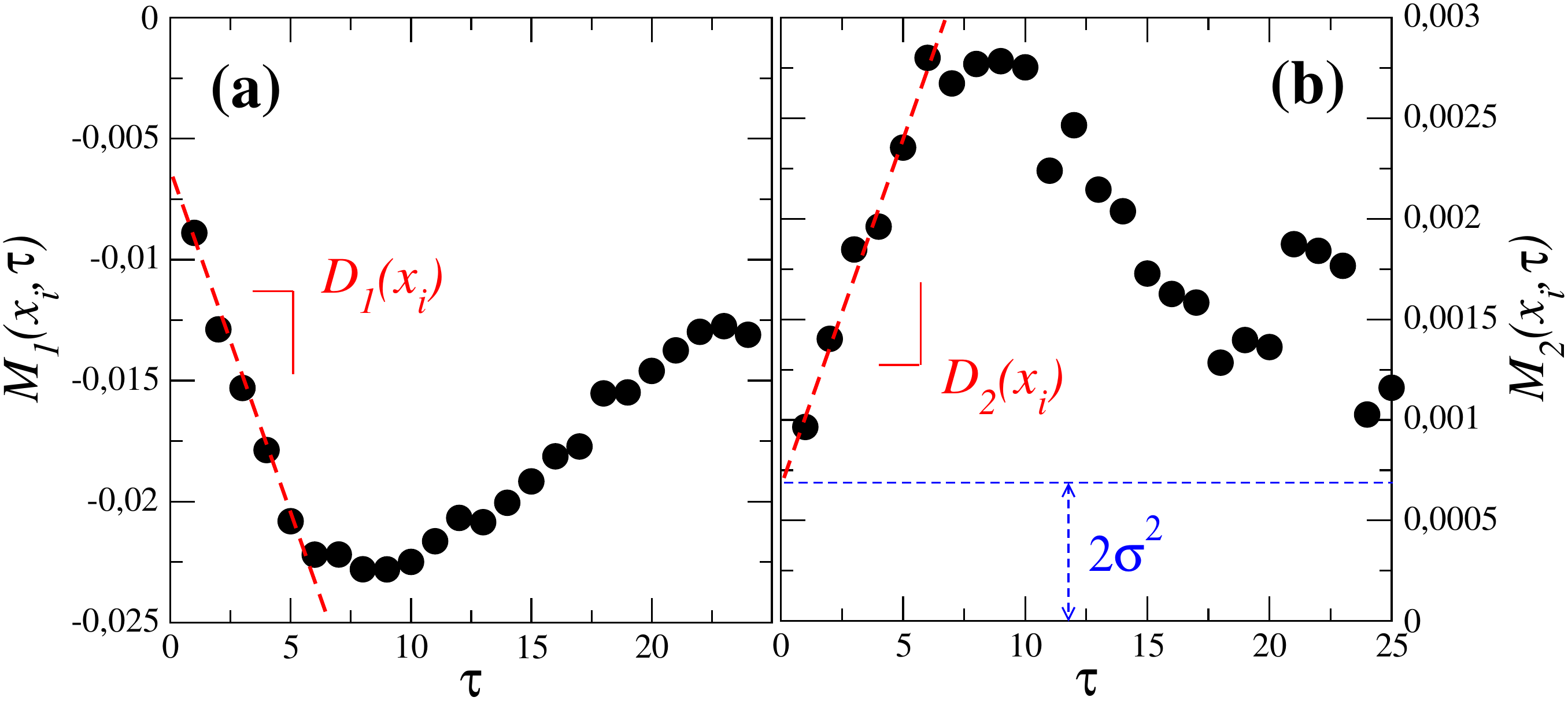}
\caption{\protect
         Illustration of the conditional moments computed directly
         from the time series of the $\phi$ time-series for
         the inverse-$\Gamma$:
         {\bf (a)} first conditional moment $M^{(1)}$ and
         {\bf (b)} first conditional moment $M^{(2)}$, from which
         one can conclude about the possible existence of
         measurement noise sources (see text). 
         Here $x_i$ is the bin including the average value 
         $\langle \phi\rangle$.}
\label{fig04}
\end{figure}

Taking the time series of the parameter $\phi$ we derive the 
stochastic evolution equation as thoroughly described in 
Ref.~\cite{friedrich11}.
This approach retrieves two functions, called the drift and
diffusion coefficients\cite{Risken}, $ D_1(\phi)$ and $D_2(\phi)$, 
governing the stochastic evolution of $\phi$:
\begin{equation}
d\phi = D_1(\phi)dt + \sqrt{D_2(\phi)}dW_t .
\label{xlangevin}
\end{equation}
Where $W_t$ represents the typical Wiener process, with $\langle W_t \rangle = 0$ and $\langle W_tW_t'\rangle=2\delta(t-t')$.
Typically the drift term governs the deterministic contributions
for the overall evolution of $\phi$, while the diffusion term
governs the corresponding (stochastic) fluctuations.

Functions $D_1(\phi)$ and $D_2(\phi)$ can be computed directly
from the data\cite{friedrich11} computing the first and second
conditional moments respectively ($n=1,2$):
\begin{equation}
D_n (\phi_i) = \lim_{\tau \to 0}\frac{1}{n!\tau} M_n(\phi_i,\tau) ,
\label{KramersMoyal}
\end{equation}
where $\phi_i$ represents one specific bin-point in the 
range of observable values and the conditional moment is given by
\begin{equation}
M_n(\phi_i,\tau) = \langle (\phi(t+\tau)-\phi(t) )^n\rangle |_{\phi(t)=\phi_i} .
\label{condmoments}
\end{equation}

Figure \ref{fig04}a and \ref{fig04}b show the first and second 
conditional moments respectively, as a function of $\tau$, for a given 
bin value $\phi_i$.
For the lowest range of $\tau$ values one sees a linear dependence
of the conditional moments, which enables to directly extract the
corresponding value of the drift and diffusion in Eq.~(\ref{KramersMoyal}).
Further, there is a clear offset in both moments, which 
indicates the presence of an additional stochastic process superimposed 
on the intrinsic stochastic dynamics, called measurement
noise\cite{boettcher06}, whose amplitude can be estimated
as $\sigma=\sqrt{M_2(\langle\phi\rangle,0)/2}$\cite{lind2010}.
See Fig.~\ref{fig04}b.
\begin{figure}[t]
\centering
\includegraphics[width=0.95\textwidth]{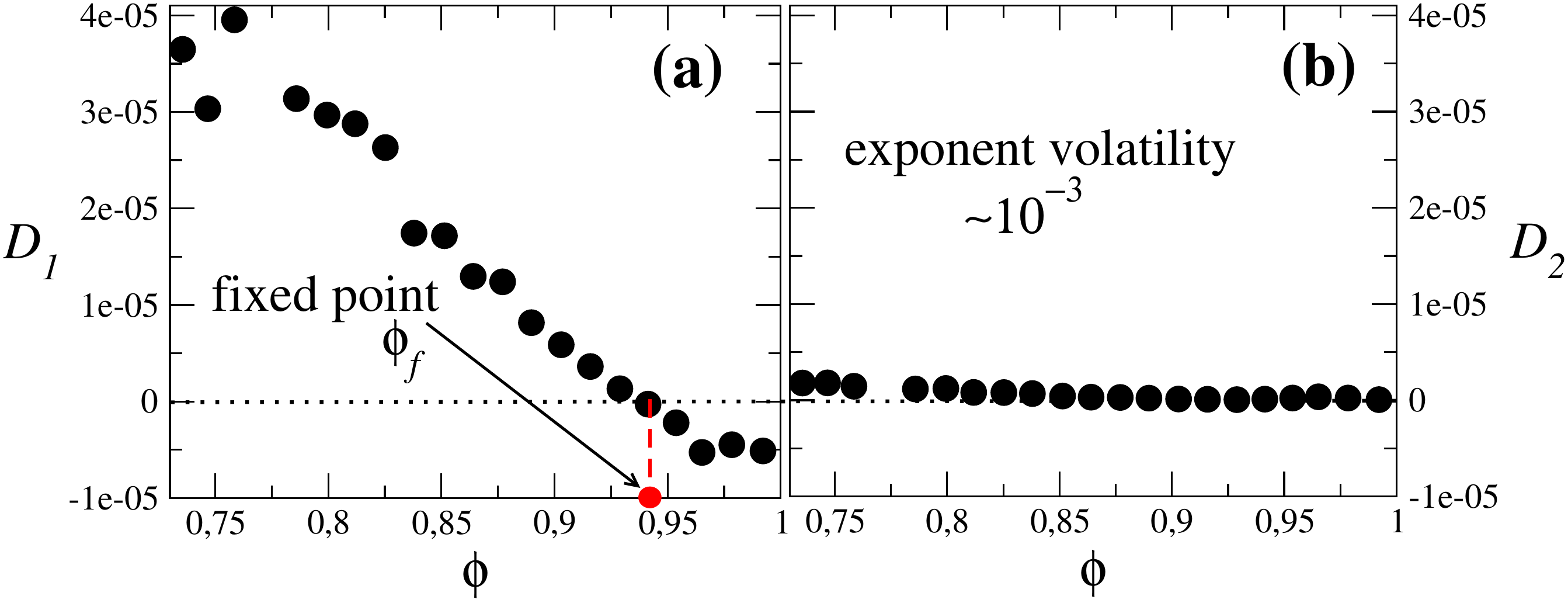}
\caption{\protect
         {\bf (a)} The drift and 
         {\bf (b)} diffusion coefficients characterizing the
         stochastic evolution of the parameter $\phi$ that describes
         the tail of the inverse-$\Gamma$ distribution (see text).}
\label{fig05}
\end{figure}

By computing the slopes of $M_1$ and $M_2$ for each bin in variable $\phi$
yields a complete definition of both drift $D_1$ and diffusion $D_2$
coefficients for the full range of observed $\phi$ values.
Figures \ref{fig05}a and \ref{fig05}b show the drift and diffusion
respectively. While the diffusion term has an almost constant amplitude,
$\sqrt{D_2}\sim 10^{-3}$, the drift is linear on $\phi$ with a negative
sloped and a fixed point close to one, $\phi_f\sim 0.93$. 

This last observation is interesting from the point of view of 
the inverse Gamma PDF: the volume-price tails fluctuate
around an inverse square law $\sim s^{-2}$ driven by a restoring 
force which can be modelled through Hooke's law. 
Furthermore, the fluctuations around the inverse square law 
are quantified by the diffusion amplitude $\sqrt{D_2}$
of the tail parameter, which can be interpreted as a sort of
``parameter volatility''.

\section{Discussion and Conclusions}
\label{sec:conclusions}

In this paper we analyse New York stock market 
volume-price distributions during the last two years
sampled every ten minutes.
We tested four models commonly applied to finance data
and presented evidence that the inverse Gamma distribution
is the model yielding the least error.

Further, we considered the parameter controlling the tail 
of the inverse Gamma distribution and extracted a Langevin
equation governing its stochastic evolution directly from
the parameter's time series.
While the deterministic contribution (drift) depends
linearly on the parameter, with a restoring force around
unity approximately, the stochastic contribution 
(diffusion) is almost constant.
Considering both contributions together, our findings
show that the tail of the volume-price distributions
tend to evolve stochastically around an inverse square law
with a constant parameter volatility.

This parameter volatility can be proposed as a risk measure
for the expected tail of New York assets.
The analysis propose here can be extended to other markets
or even in other contexts where non-stationary processes
are observed. If the inverse Gamma distribution is commonly
the best model for volume-price distributions is up to 
our knowledge an open question.
The confidence of each model can be further tested using other
methods such as the Kolmogorov-Smirnov test\cite{Kleinhans12}.

It must be noticed that the above approach is only valid 
for Markovian processes, which seems to be the case of the
parameter here considered, which was tested
comparing two-point and three-point conditional probabilities.
Moreover, 
the Langevin analysis here proposed can also be extended
to both parameters characterizing the inverse Gamma
model. 
Further research will be necessary to access  the reliability 
of the stochastic reconstruction of the volume-price evolution, 
and a comparison to theoretical agent models.
These and other issues will be addressed elsewhere.

\section*{Acknowledgments}

The authors thank 
Funda\c{c}\~ao para a Ci\^encia e a Tecnologia 
for financial support 
under PEst-OE/FIS/UI0618/2011, PEst-OE/MAT/UI0152/2011, 
F\-C\-O\-M\-P-01-0124-FEDER-016080 and SFRH/BPD/65427/2009 (FR).
This work is part of a bilateral cooperation DRI/DAAD/1208/2013 
supported by FCT and Deutscher Akademischer Auslandsdienst (DAAD).


\end{document}